\documentclass[aps,prd,superscriptaddress,showpacs,preprint]{revtex4}
\usepackage{graphicx, bm}
\usepackage[usenames]{color}

\usepackage{multirow}
\usepackage{bigstrut}
\usepackage{slashbox}

\begin{document}
\draft
\title{Higgs Boson Self-Coupling at High Energy $\gamma \gamma$ Collider}

\author{ A. Guti\'errez-Rodr\'{\i}guez}
\affiliation{\small Facultad de F\'{\i}sica, Universidad Aut\'onoma de Zacatecas\\
         Apartado Postal C-580, 98060 Zacatecas, M\'exico.\\}

\author{Javier Peressutti}
\affiliation{\small Instituto de F\'{\i}sica de Mar del Plata
                    (IFIMAR) CONICET, UNMDP.
                    Departamento de F\'{\i}sica, Universidad Nacional del Mar del Plata\\
                    Funes 3350, (7600) Mar del Plata, Argentina.}

\author{O. A. Sampayo}
\affiliation{\small Instituto de F\'{\i}sica de Mar del Plata
                    (IFIMAR) CONICET, UNMDP.
                    Departamento de F\'{\i}sica, Universidad Nacional del Mar del Plata\\
                    Funes 3350, (7600) Mar del Plata, Argentina.}

\date{\today}

\begin{abstract}

We analyzed the double production and the triple self-coupling of
the standard model Higgs boson at future $\gamma \gamma$ collider
energies, with the reactions $\gamma\gamma \rightarrow f \bar f
HH$ $(f=b, t)$. We evaluated the total cross section for $f\bar
fHH$ and calculated the total number of events considering the
complete set of Feynman diagrams at tree-level and for different
values of the triple coupling $\kappa\lambda_{HHH}$. We have also
analyzed the sensitivity for the considered reaction and we show
the results as 95\% C.L. regions in the $\kappa-M_H$ plane for
different values of the center of mass energy and different levels
of background. The numerical computation was done for the energies
which are expected to be available at a possible Future Linear
$\gamma\gamma$ Collider with a center-of-mass energy $500-3000$
$GeV$ and luminosities of 1 and $5\hspace{0.5mm} ab^{-1}$. We
found that the number of events for the process $\gamma\gamma
\rightarrow t \bar t HH$, taking into account the decay products
of both $t$ and $H$, is small but enough to obtain information on
the triple Higgs boson self-coupling in a independent way,
complementing other studies on the triple vertex.

\end{abstract}

\pacs{ 13.85.Lg, 14.80.Bn
\vspace*{2cm}\noindent  E-mail: $^{1}$alexgu@fisica.uaz.edu.mx,
$^{2}$sampayo@mdp.edu.ar}

\vspace{5mm}

\maketitle

\section{Introduction}

The Higgs boson \cite{Higgs} plays an important role in the
Standard Model (SM) \cite{SM} because it is responsible for
generating the masses of all the elementary particles (leptons,
quarks, and gauge bosons). However, the Higgs-boson sector is the
least tested in the SM, in particular the Higgs boson
self-interaction.

The search for Higgs bosons is one of the principal missions of
present and future high-energy colliders. The observation of this
particle is of major importance for the present understanding of
fundamental interactions. Indeed, in order to accommodate the well
established electromagnetic and weak interaction phenomena, the
existence of at least one isodoblete scalar field to generate
fermion and weak gauge bosons masses is required. Despite repeated
success in explaining the present data, the SM cannot be
completely tested before this particle has been experimentally
observed and its fundamental properties studied.

The triple and quartic Higgs boson couplings
\cite{Boudjema,Ilyin,Djouadi} $\lambda$ and $\tilde\lambda$ are
defined through the potential

\begin{equation}
V(\eta_H)=\frac{1}{2}M^2_{H}\eta^2_{H}+\lambda
v\eta^3_{H}+\frac{1}{4}\tilde\lambda\eta^4_{H},
\end{equation}

\noindent where $\eta_{H}$ is the physical Higgs field. In the SM,
we obtain $M_H= \sqrt{2\lambda}v$ as the simple relationship
between the Higgs boson mass $M_{H}$ and the self-coupling
$\lambda$ where $v=246$ $GeV$ is the vacuum expectation value of
the Higgs boson. The triple vertex of the Higgs field $H$ is given
by the coefficient $\lambda_{HHH}=\frac{3M^2_H}{M^2_Z}$. An
accurate test of this relationship may reveal the extended nature
of the Higgs sector. The measurement of the triple Higgs boson
coupling is one of the most important goals of Higgs physics in a
future $e^+e^-$ linear collider experiment. This would provide the
first direct information on the Higgs potential that is
responsible for electroweak symmetry breaking.

The future $e^+e^-$ linear collider can not only be designed to
operate in $e^+e^-$ collision mode, but can also be operated as a
$\gamma\gamma$ collider. This is achieved by using Compton
backscattered photons in the scattering of intense laser photons
on the initial $e^+e^-$ beams. The design of the photon linear
collider and physics opportunities are described in references
\cite{ILC,Badelek} and the possibility of measuring the $HHH$
coupling is discussed in Refs. \cite{Ilyin,Jikia,Belusevic}.

The triple Higgs boson self-coupling can be measured directly in
pair-production of Higgs particles at hadron and high-energy
$e^+e^-$ linear colliders. Several mechanisms that are sensitive
to $\lambda_{HHH}$ can be exploited for this task. Higgs pairs can
be produced through double Higgs-strahlung of $W$ or $Z$ bosons
\cite{Djouadi,Gounaris,Kamoshita}, $WW$ or $ZZ$ fusion
\cite{Ilyin,Barger,Dobrovolskaya,Dicus} as well as through
gluon-gluon fusion in $pp$ collisions \cite{Glover,Plehn,Dawson}
and high-energy $\gamma\gamma$ fusion \cite{Ilyin,Jikia} at photon
colliders. In $\gamma\gamma$ collisions, double Higgs production
is possible in several reactions at the tree level, and one
process, $\gamma\gamma \to HH$, at the one-loop level:

\begin{eqnarray}
\mbox{$WW$ double Higgs-fusion}&:& \gamma\gamma \to WWHH,  \nonumber \\
\mbox{double Higgs-fermion}&:& \gamma\gamma \to f\bar f HH,\\
\mbox{one-loop level}&:& \gamma\gamma \to HH.\nonumber
\end{eqnarray}

As was studied in \cite{Ilyin} the reaction $\gamma\gamma \to
WWHH$ is free from any background except incorrect combinatorial
of jets; for $M_H < 150$ $GeV$, the main branching is four
$b$-jets (with an invariant mass peak at $M_H$) plus jets from two
additional $W$ bosons, up to four, and/or a large missing energy.
8-jet events can be detected in total: For $M_H > 150$ $GeV$,  the
signature includes up to twelve quark jets with the invariant mass
peaking at $M_W$ and $M_H$.

In the case of the process $\gamma\gamma \to f\bar f HH$, the
important contribution happens when the considered fermion is a
quark-top since the Higgs-fermion coupling is proportional to the
fermion mass. For completeness, we have also calculated the
contribution when the fermion is a $b$-quark. We have found that
this last contribution is negligible $(\sigma\sim 10^{-6} $ $fb$
for $M_H=120$ $GeV$ and $\kappa=2$).

The study of the four-body processes with quark top $t$,
$\gamma\gamma\rightarrow t\bar t HH$, in which the SM Higgs boson
is radiated by a $t(\bar t)$ quark at future $\gamma\gamma$
colliders \cite{ILC,Badelek} with a c.m. energy in the range of
500 to 3000 $GeV$, as in the case of the DESY TeV Energy
Superconducting Linear Accelerator (TESLA) machine \cite{TESLA},
is necessary in order to know its impact on other processes and
also to search for new relations that could have a clear signature
of the Higgs boson production.

Another process with double Higgs production in $\gamma\gamma$
collisions is $\gamma\gamma \to HH$. This reaction proceeds at the
one-loop level. Analytical results for the amplitude and a
detailed numerical analysis were carried out in Refs. \cite{Jikia,
Belusevic}.

The Higgs coupling with top quarks, the largest coupling in the
SM, is directly accessible in the process where the Higgs boson is
radiated off top quarks $\gamma\gamma\rightarrow t\bar t HH$.
Consequently, this process can be used to probe the $t-\bar t-H$
Yukawa coupling directly. This process also depends on the Higgs
boson triple self-coupling, which could lead us to obtain the
first non-trivial information on the Higgs potential. We are
interested in finding regions that could allow the observation of
the $t\bar tHH$ processes at the next generation of high energy
$\gamma\gamma$ linear colliders. We consider the complete set of
Feynman diagrams at tree-level (Fig. 1).

This paper is organized as follows: In Sec. II, we study the
triple Higgs boson self-coupling through the processes
$\gamma\gamma \rightarrow f \bar f HH$ at next generation linear
$\gamma\gamma$ colliders. In Sec. III, we have considered the
background for the process $\gamma\gamma \rightarrow t \bar t HH$
and we have studied their sensitivity to different values of the
triple vertex. Finally, we summarize our results in Sec. IV.

\section{Cross-Section of the Higgs Boson Double Production with Triple
Self-Coupling}

In this section we present numerical results for $\gamma\gamma
\rightarrow f \bar f HH$ $(f=b, t)$ with double Higgs boson
production. Since the Higgs-fermion coupling is proportional to
the fermion mass, we analyzed the $t$-quark case in more detail.
We carried out the calculations using the framework of the
Standard Model at next generation linear $\gamma\gamma$ colliders.
For the initial photon we use the spectrum of backscattered
photons for unpolarized beams as given in Ref. \cite{Ginzburg}.

We use the CALCHEP \cite{Pukhov} packages to check the different
parts of the calculations of the matrix elements and cross-sections.
These packages provide automatic computation of the cross-sections
and distributions in the SM as well as their extensions at
tree-level. We consider the high energy stage of a possible Next
Linear $\gamma\gamma$ collider with $\sqrt{s}=500-3000$ $GeV$ and
design luminosity 1 and $5\hspace{0.5mm}ab^{-1}$.

\subsection{Triple Higgs Boson Self-Coupling Via $\gamma\gamma \rightarrow f \bar f HH$ $(f=b, t)$}

To illustrate our results of the sensitivity to the $HHH$ triple
Higgs boson self-coupling, we show the dependence of the
cross-section on the center-of-mass energy of $\sqrt{s}=500-3000$
$GeV$ for $\gamma\gamma \rightarrow t \bar t HH$ in Fig. 2 for
several values of the Higgs boson mass $M_H= 120, 140$ $GeV$. The
variation of the cross-section for the modified triple couplings
$\kappa\lambda_{HHH}$ is evaluated for some values of $\kappa$ in
the range ($-2,2$). The cross-section is sensitive to the value of
the triple couplings as well as to the Higgs boson mass. The
sensitivity to $\lambda_{HHH}$ increases with the collider energy,
reaching a maximum at the end of the range considered:
$\sqrt{s}\sim 3000$ $GeV$. As an indicator of the order of
magnitude we present the Higgs boson number of events in Table 1
(of course we have to multiplicate for the corresponding Branching
Ratios to obtain the observable number of events) for several
Higgs boson masses, center-of-mass energy and $\kappa$ values and
for a luminosity of $1\hspace{0.5mm}ab^{-1}$ and $5\hspace{0.5mm}
ab^{-1}$.

We also include a contours plot for the number of events of the
studied processes in the $(\sqrt{s}, \kappa)$ plane with $M_H=
120,140$ $GeV$ and $5\hspace{0.5mm} ab^{-1}$ in Figs. 3 and 4.
These contours are obtained from Table 1.

Finally, for the $\gamma\gamma \rightarrow b \bar b HH$ process,
the cross-section as a function of the center-of-mass energy
$\sqrt{s}$ for $M_H=120$ $GeV$ and $\kappa=2$ is negligible
($\sigma \sim 10^{-6} fb$). In these conditions we did not have
the possibility to detect this process.

\vspace{5mm}

\begin{center}
\begin{tabular}{|c|c|c|c|c|c|c|c|c|c|c|}
\hline \multirow{2}{*}{}{} $\gamma\gamma \rightarrow t \bar tHH$ &
\multicolumn{5}{|c|}{$M_H=120$ $GeV$} &
\multicolumn{5}{|c|}{$M_H=140$ $GeV$}\\
\hline \hline
\backslashbox{$\sqrt{s} (GeV)$}{$\kappa$}&  -2 & -1 & 0 & 1 & 2 & -2 & -1 & 0 & 1 & 2 \\
\hline
\hline

800   & - (-) & - (-) & - (-) & - (-) & - (-) & - (-) & - (-) & - (-) & - (-) & - (-) \\
\hline

1000   & - (-) & - (-) & - (-) & - (-) & - (-) & - (-) & - (-) & - (-) & - (-) & - (-) \\
\hline

1500   & 2 (10) & 3 (15) & 3 (15) & 4 (20) & 4 (20) & 1 (5) & 1 (5) & 1 (5) & 2 (10) & 2 (10) \\
\hline

2000  & 5 (25) & 5 (25) & 5 (25) & 6 (30) & 7 (35) & 3 (15) & 3 (15) & 3 (15) & 3 (15) & 4 (20) \\
\hline

2500  & 6 (30) & 6 (30) & 6 (30) & 8 (40) & 9 (45) & 4 (20) & 4 (20) & 4 (20) & 4 (20) & 6 (30) \\
\hline

3000  & 7 (35) & 7 (35) & 7 (35) & 8 (40) & 10 (50)& 5 (25) & 5 (25) & 5 (25) & 5 (25) & 7 (35) \\
\hline
\end{tabular}
\end{center}

\begin{center}
Table 1. Total production of Higgs boson pairs in the SM for
${\cal L}=1\hspace{1mm} (5)\hspace{0.5mm}ab^{-1}$ and $\kappa=-2,
-1, 0, 1, 2$.
\end{center}

\section{Backgrounds and Sensitivity}

For the process that we are studying, $\gamma \gamma \rightarrow t
\bar{t} H H$, we have considered the following background: $\gamma
\gamma \rightarrow t \bar{t} Z H$ and $\gamma \gamma \rightarrow t
\bar{t} Z Z$. If we consider a center-of-mass energy of
$\sqrt{s}=2000$ $GeV$ of the parent $e^+e^-$ and a Higgs boson
mass of $M_H=120$ $GeV$ these process have total cross section of
0.006 $fb$ and 0.004 $fb$ respectively, while the signal process
has 0.006 $fb$. Considering that $Br(H\rightarrow b\bar{b})\simeq
0.7$ and $Br(Z\rightarrow b\bar{b})\simeq 0.15$ and after
$b$-tagging one can expect the ratio to be $S/B\sim 4.5$. In Fig.
5 (a) we show the ratio

\begin{equation}
\frac{S}{B} = \frac{\sigma_{t \bar{t} H H} Br(H\rightarrow
b\bar{b}) Br(H\rightarrow b\bar{b})} {\sigma_{t \bar{t} Z H}
Br(H\rightarrow b\bar{b})Br(Z\rightarrow b\bar{b})+\sigma_{t
\bar{t} Z Z} Br(Z\rightarrow b\bar{b}) Br(Z\rightarrow b\bar{b})
},
\end{equation}

\vspace{2mm}

\noindent as a function of the center-of-mass energy $\sqrt{s}$
and different values of the Higgs boson mass. We also consider the
separate contributions of the $t\bar t HZ$ and $t\bar t ZZ$
backgrounds in Fig. 5 (b).

We should note that the $b$-tagging would remove the possible
background from $W$ bosons due to a small branching ratio of
$W\rightarrow cs$ and then we neglect the background from $W$
bosons that come from the $\gamma \gamma \rightarrow b \bar{b} W^+
W^-$ process.

In order to quantify the sensitivity of the considered reaction to
different values of $\kappa$ we use the statistical sensitivity,
$S_{stat}$, which is  defined as \cite{Takahashi}:

\begin{equation}
S_{stat}\equiv\frac{
N(\kappa)-N_{SM}}{\sqrt{N_{obs}}}=\frac{L_{tot}\mid
\sigma(\kappa)-\sigma_{SM}
\mid}{\sqrt{L_{tot}[\sigma(\kappa)+\eta_B \sigma_B]}},
\end{equation}

\noindent where, $N(\kappa)$ is the expected number of events as a
function of $\kappa$, $N_{SM}$ is the number of events expected
from the standard model and $N_{obs}$ is the number of observed
events. While, $L_{tot}, \sigma(\kappa), \sigma_{SM}, \eta_B$ and
$\sigma_B$ are the total integrated luminosity, the cross section
as a function of $\kappa$, the standard model cross section
($\kappa=1$), the intensity level for background events and the
cross section for the background process, respectively.

In Fig. 6 we show the statistical analysis when the experiment
does not show any deviation from the SM at 95\% C.L. for the
$\lambda_{HHH}$ coupling. These regions correspond to $S < 1.96$
(95\% C.L. correspond to $S^2=\chi^2<(1.96)^2$) for differents
values of the center-of-mass energy, the $e^+e^-$ luminosity and
the parameter $\eta_B$.

The quantity $\eta_B$ measure the background importance. As the
masses we are considering for the Higgs are different of the Z
boson mass then it could be possible separate the Higgs signal
from the Z background by reconstructing the jet-jet mass. In this
conditions we also take into account the possibility of a small
background. In Fig. 6 we show the case with very small background
($\eta_B=0$) and full background ($\eta_B=1$).

\section{Conclusions}

As a possible option of the International Linear Collider (ILC),
the feasibility of physics opportunities of high energy physics
photon-photon interaction has been considered in Ref. \cite{Boos}.
In the high energy photon linear collider, high energy photon
beams are generated by inverse Compton scattering between the
electron and the laser beams. The $\gamma\gamma$ collider
represents a possible opportunity for the triple Higgs boson
self-coupling analysis. We have analyzed the triple Higgs boson
self-coupling at future $\gamma\gamma$ collider energies with the
reactions $\gamma\gamma \rightarrow f \bar f HH$ where $f=b, t$
and considering the complete set of Feynman diagrams at tree-level
and in the frame work of the standard model.

We found that for the process $\gamma\gamma \rightarrow t \bar t
HH$, the complete calculation at tree level gives a production
cross-section of the order of a fraction of a femtobarn, i.e.,
$0.010$ $fb$ and $0.007$ $fb$ for $M_H=120, 140$ $GeV$,
respectively, and at the end of the examined energy range. These
values are, however, larger than the production cross-section for
$e^{+}e^{-}\rightarrow b \bar b HH$ and $e^{+}e^{-}\rightarrow t
\bar t HH$ \cite{A.Gutierrez}. The number of events obtained
considering the decay products of both $t$ and $H$ is enough to
obtain relevant information about the triple Higgs self-coupling.
Moreover, this process can be used to probe the $t-\bar t-H$
Yukawa coupling.

We have also done a background analysis concluding that the
background intensity is below the signal level. The ratio
$S\mbox{(signal)}/B\mbox{(background)}$ is show in Fig. 5. On the
other hand we have also studied the sensitivity to the values of
the parameter $\kappa$. We show the results as 95\% C.L. regions
in the $\kappa-M_H$ plane for different values of the
center-of-mass energy $\sqrt{s}$, the luminosity $\cal{L}$ and the
parameter $\eta_B$. These are shown in Fig. 6. The limits for
$\kappa$ are $M_H$ dependent and we can see that the most
stringent limit are possible for positive values of $\kappa$.
Finally, the study of this process is important and could be
useful to probe the triple Higgs boson self-couplings
$\lambda_{HHH}$ using a $\gamma \gamma$ collider with very high
luminosity and high center-of-mass energy. In addition, to our
knowledge,  these results have never been reported in the
literature before and could be of relevance for the scientific
community since the studied process is complementary of other
studied in the literature.

\vspace{1.5cm}

\begin{center}
{\bf Acknowledgments}
\end{center}

We acknowledge support from CONACyT, SNI and PROMEP (M\'exico). O.
A. Sampayo would like to thank CONICET (Argentina).

\newpage

\begin{figure}[t]
\centerline{\scalebox{1.0}{\includegraphics{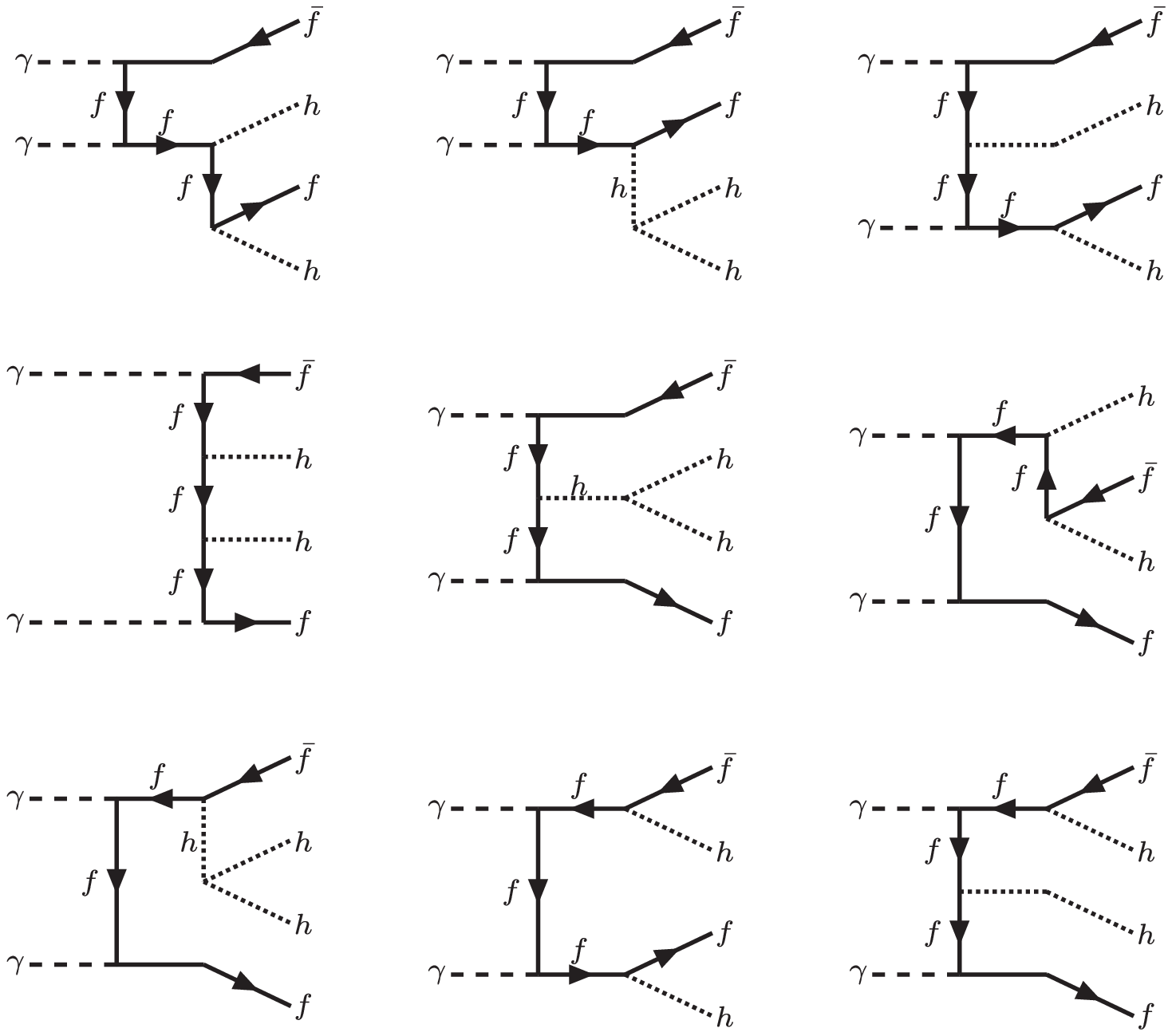}}}
\caption{ \label{fig:gamma} Feynman diagrams at tree-level (plus
fermion inverted circulation) for $\gamma\gamma \rightarrow f\bar
f HH$ with $f=b, t$.}
\end{figure}

\begin{figure}[t]
\centerline{\scalebox{1.0}{\includegraphics{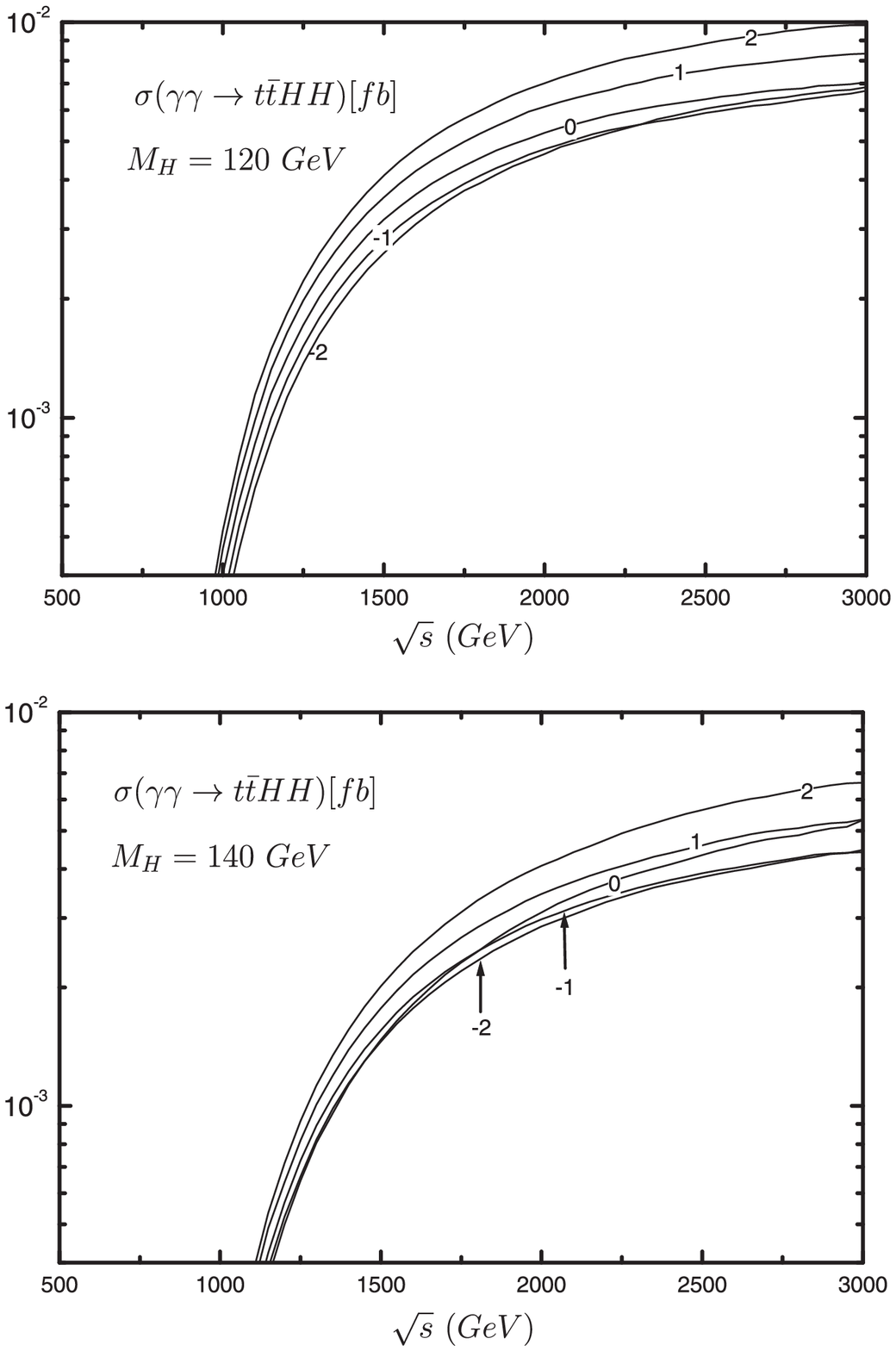}}}
\caption{ \label{fig:gamma} The dependence of the cross-section
with the center-of-mass energy $\sqrt{s}$ for two fixed Higgs
masses $M_H= 120, 140$ $GeV$. The variation of the cross-section
for modified triple couplings $\kappa \lambda_{HHH}$ is indicated
by $\kappa=-2, -1, 0, 1, 2$.}
\end{figure}

\begin{figure}[t]
\centerline{\scalebox{0.9}{\includegraphics{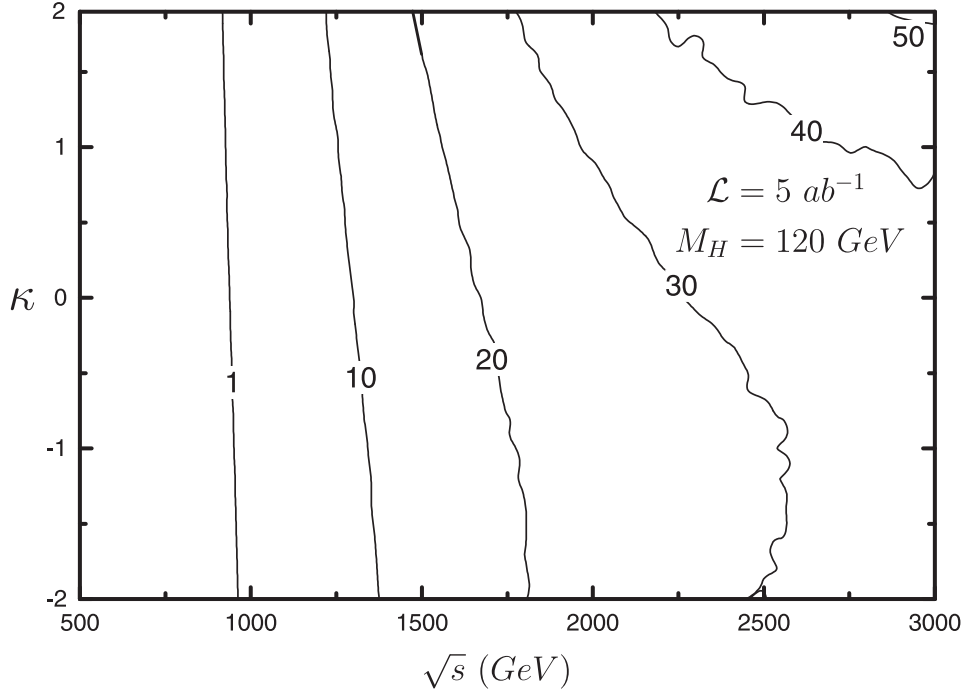}}}
\caption{ \label{fig:gamma} Contour plot for the number of events
of the process $e^{+}e^{-}\rightarrow t\bar t HH $ as a function
of $\sqrt{s}$ and $\kappa$. The variation of the number of events
for modified triple couplings $\kappa\lambda_{HHH}$ is indicated
for ${\cal L}=5$ $ab^{-1}$ and $M_H= 120$ $GeV$.}
\end{figure}

\begin{figure}[t]
\centerline{\scalebox{0.9}{\includegraphics{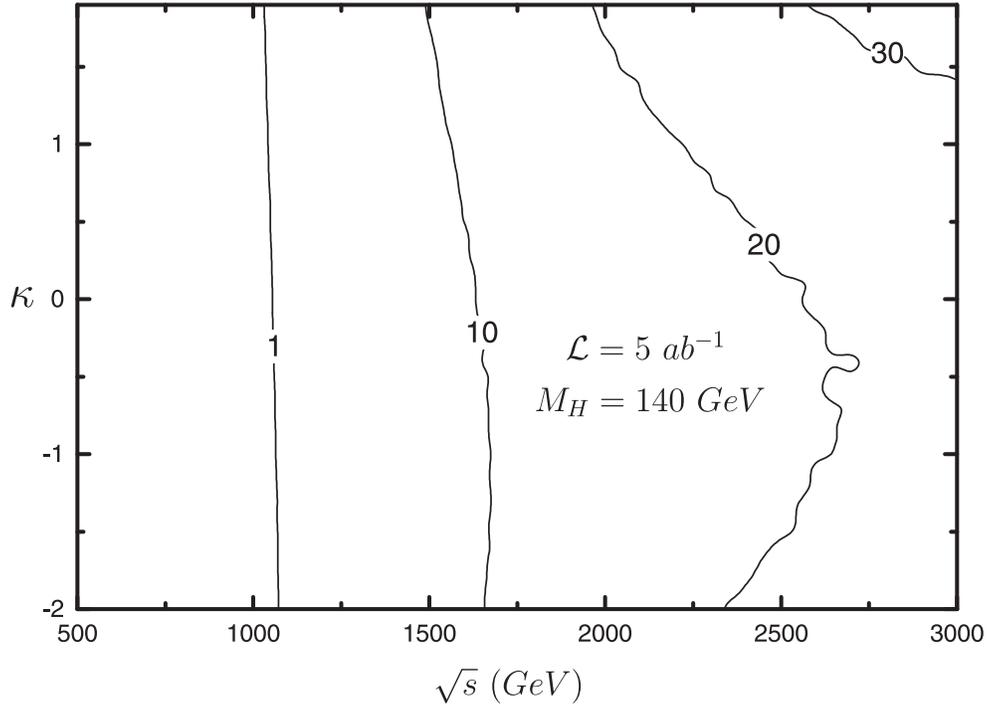}}}
\caption{ \label{fig:gamma} The same as in Fig. 3 but for $M_H=
140$ $GeV$.}
\end{figure}

\begin{figure}[t]
\centerline{\scalebox{1.0}{\includegraphics{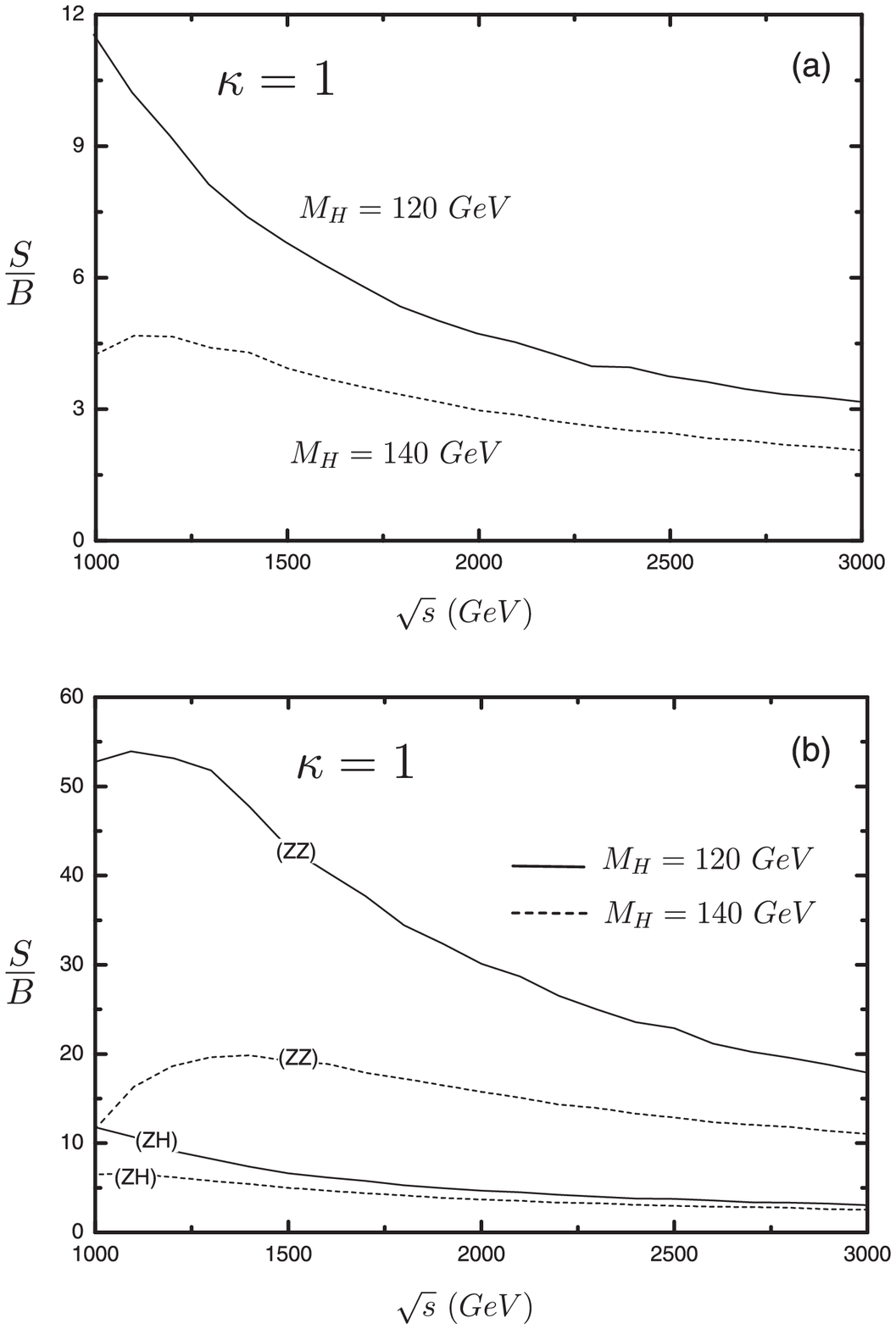}}}
\caption{ \label{fig:gamma} (a) Ratio S(signal)/B(background) as a
function of center-of-mass energy to the process $\gamma\gamma
\rightarrow t\bar t HH$ with $M_H=120, 140$ $GeV$ and $\kappa=1$.
(b) Ratio $S/B$ for the separate  contributions of the $t\bar tHZ$
and $t\bar tZZ$ backgrounds with $M_H=120, 140$ $GeV$ and
$\kappa=1$.}
\end{figure}

\begin{figure}[t]
\centerline{\scalebox{1.0}{\includegraphics{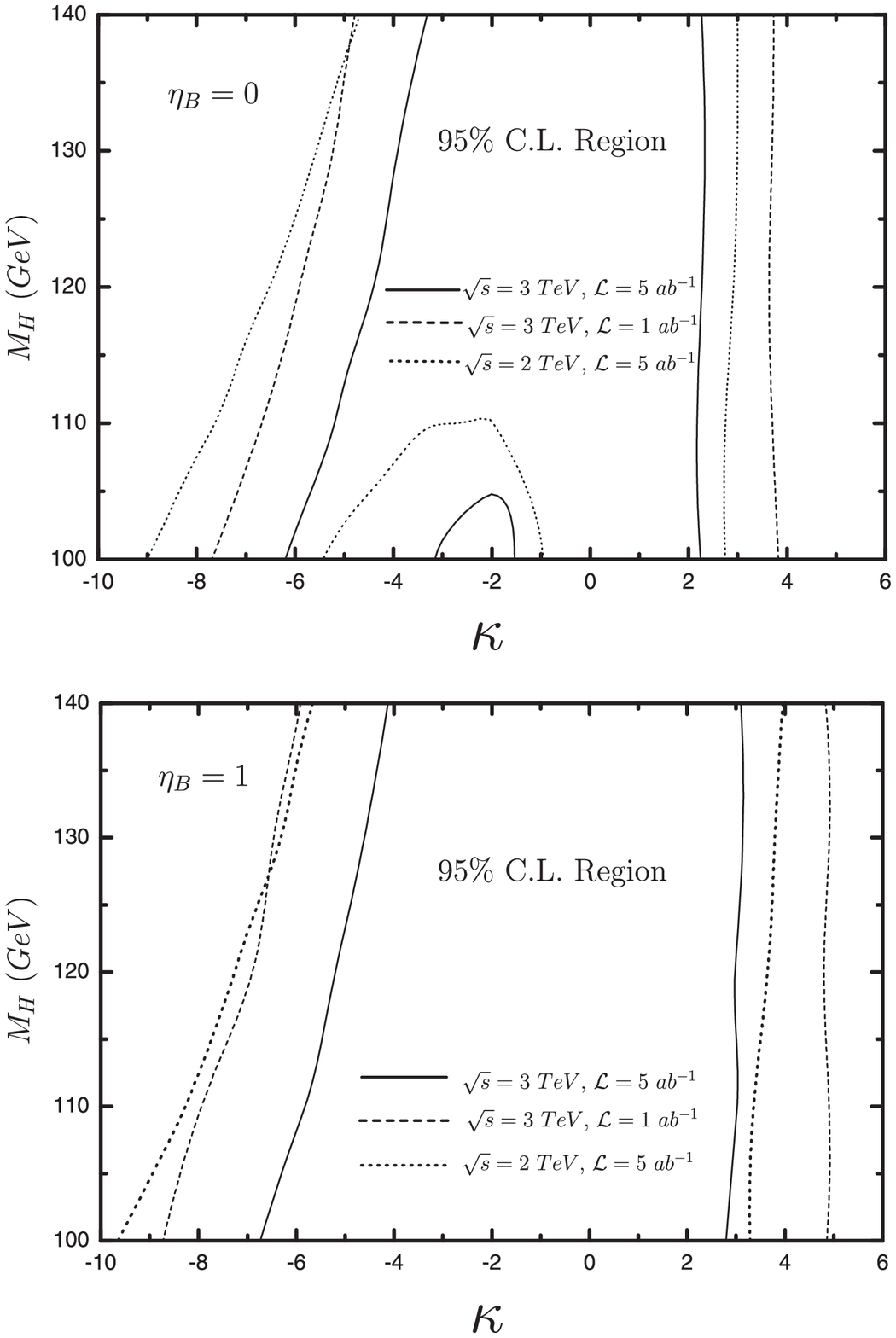}}}
\caption{ \label{fig:gamma} {Region in the ($\kappa-M_H$) plane
where the experiment does not show any deviation from the SM
($\kappa=1$) at 95\% C.L. for $\sqrt{s}=2, 3$ $TeV$, ${\cal L}=1,
5$ $ab^{-1}$ and $\eta_B=0, 1$.}}
\end{figure}

\end{document}